\documentclass[paper]{JHEP3}

\def\psibar{\bar{\psi}}
\def\lambdabar{\bar{\lambda}}
\def\zetabar{\bar{\zeta}}
\def\bea{\begin{eqnarray}}
\def\eea{\end{eqnarray}}
\def\be{\begin{equation}}
\def\ee{\end{equation}}
\def\Bp2{B(p^2)}
\def\Bq2{B(q^2)}
\def\Ap2{A(p^2)}
\def\Aq2{A(q^2)}
\def\Tr{\mbox{Tr}}
\def\half{\frac{1}{2}}

\title{Three point SUSY Ward identities without Ghosts}
\author{M.L.Walker}
\abstract{We utilise a non-local gauge transform which renders the entire
action of SUSY QED invariant and respects the SUSY algebra 
modulo the gauge-fixing condition, to derive two- and three-point
ghost-free
SUSY Ward identities in SUSY QED. We
use the cluster decomposition principle to find the Green's function Ward 
identities and then takes linear combinations of the latter to derive 
identities for the proper functions.}

\preprint{hep-th/0408134}

\begin{document}

\section{Introduction}
Supersymmetry (SUSY) is a well-established theory expected to play an essential
role in any non-trivial unification of gravity with the gauge forces. More
immediate concerns are the solution to the heirachy problem and physics 
immediately beyond the standard model. 

It has long been known that analysis of gauge field theories can
be greatly simplified by exploiting the gauge symmetry to derive 
Ward identities that relate the various 
Green's functions to each other~\cite{W50}. The same is even more true of 
Ward-Takahashi identities, 
which relate the proper functions of a gauge theory~\cite{BC80}.

The application of this concept to SUSY is almost as hold as SUSY itself,
with identities relating the propagators of fields within the same 
multiplet~\cite{IZ74}. 
The application to SUSY gauge theories followed soon with the work of 
Piquet and Sibold~\cite{PS86} and others~\cite{RSS01,OR02}, who dealt with
the complications due to auxiliary fields, Wess-Zumino gauge and gauge fixing,
using the BRST quantization to derive SUSY Slavnov-Taylor
identities. While these are 
long established procedures of gauge field theory in the context of gauge 
fixing, their application to SUSY is not entirely analogous. Ghosts are 
necessary in non-Abelian gauge theories as part of the gauge-fixing
procedure, without which the gauge boson does not 
have a well-defined propagator. However these considerations do not
apply to SUSY. If one is content to simply do perturbative calculations then
all propagators are well-defined, once the gauge is fixed, and further ghosts
are not necessary. 

This is not to say that such methods are not valuable.
Indeed, they have allowed the development of renormalisation schemes that
respect both SUSY and gauge symmetry~\cite{S02,SS02,SS03}. 
Still, the ghosts are not inherent to SUSY in the
way that they are to gauge theory, even in Wess-Zumino gauge. 
Furthermore, this now standard technique
requires the introduction of ghosts even to the Abelian SUSY QED~\cite{OR02}, 
where there were none in the non-SUSY case.

The complications related to the gauge-fixing term in deriving SUSY identities
can be traced back to the use of the transforms derived by Wess and Zumino~\cite{WZ74}.
When working in the Wess-Zumino gauge, in which the gauge dependant 
superpartners of the photon have been gauged away, their transform leaves
the gauge invariant part of the SUSY QED action invariant but spoils the
$U(1)$ gauge fixing condition. It is at this point that standard treatments
turn to the BRST formalism. We demonstrated an alternative approach 
in \cite{W04a}. Our approach was to treat the breaking of the 
$U(1)$ gauge condition in the same way that Wess and Zumino treated the
breaking of the Wess-Zumino (WZ) gauge, by introducing another gauge transformation
to restore it. The resulting transform had the attractive feature of obeying 
the SUSY algebra modulo the gauge condition. We went on in that paper to
find SUSY Ward identities relating the propagators not only 
of the matter multiplet, but also of the gauge multiplet. These identities
all held exactly at tree level, something that had not been achieved with the 
gauge multiplet before without using 
ghost fields. Using the cluster decomposition
principle~\cite{W96,M93} to handle the non-linearities, as demonstrated by
\cite{OR02}, we then went on to find an
identity relating some three-point Green's functions.

Another important incentive for introducing ghosts is that the Wess-Zumino 
transforms, as well as our own, are non-linear. Indeed, the additional 
correction found in \cite{W04a} is also non-local. This forbids the use
of the Legendre transform or the effective action in any way, unless the 
transform is linearised as the BRST transforms are. We shall circumvent this
problem by taking linear combinations of the Green's function Ward identities.

In this paper we continue our previous work, begining with a brief 
summary of our transform and its properties in section~\ref{sec:WZL}.
In section~\ref{sec:Greens} we give a detailed discussion of identities 
relating three-point Green's functions and present them in their 
entirety in appendix~\ref{app:Greens}. In section~\ref{sec:vertex} we tackle 
the problem of SUSY Ward identities relating the vertices. Effective 
action techniques cannot be used for reasons already noted. The vertex 
identities themselves are given in appendix \ref{app:vertex}.
It shall transpire that the vertex Ward identities have already been found
by an independant method~\cite{WB99b}, 
and that their solution, when subject to reasonable
constraints like gauge invariance, has already been found in its most general 
form which we present in appendix~\ref{app:solution}.

\section{The Wess-Zumino-Lorentz Transform} \label{sec:WZL}
Since the transformations used in this paper are newly-discovered~\cite{W04a},
we briefly restate them here. The basic problem is to modify the 
identities found by
Wess and Zumino to leave the gauge invariant part of the action 
invariant~\cite{WZ74} so that they also leave the gauge-fixing component 
invariant. Their transforms are
\begin{eqnarray}
\delta_{WZ} a &=& -i\bar{\zeta} \psi \nonumber \\
\delta_{WZ} b &=& \bar{\zeta} \gamma_5 \psi \nonumber \\
\delta_{WZ} \psi &=& (f+i\gamma_5 g)\zeta 
+ i\not \! \partial (a+i\gamma_5 b) \zeta 
-e \not \! A (a-i\gamma_5 b)\zeta\nonumber \\
\delta_{WZ} f &=& \bar{\zeta} \not\!\partial \psi 
-e\bar{\zeta} [a\lambda +i b \gamma_5 \lambda 
-i\not \! A \psi ] \nonumber \\
\delta_{WZ} g &=& i\bar{\zeta} \gamma_5 \not\!\partial \psi
-ei\bar{\zeta} [a\gamma_5 \lambda +i b \lambda 
+i \not \! A \gamma_5 \psi] , \\
\delta_{WZ} A_\mu & = & \bar{\zeta} \gamma_\mu \lambda - i\bar{\zeta} \partial_\mu
\chi \nonumber \\
\delta_{WZ} \lambda & = & \frac{1}{2} 
(\gamma^\nu \gamma^\mu - \gamma^\mu \gamma^\nu )
\partial_\mu A_\nu \zeta + i\gamma_5 D \zeta \nonumber \\
\delta_{WZ} D & = & i\bar{\zeta} \gamma_5 \! \not\!\partial \lambda.
\end{eqnarray}
The effect of $\delta_{WZ}$ on the Lorentz gauge condition is
\begin{equation}
\delta_{WZ} (\partial \cdot A) = \bar{\zeta} \not \! \partial \lambda,
\end{equation}
but this can be undone with the gauge transformation
\begin{equation}
\delta_L A_\mu = -\partial_\mu \frac{\not \! \partial}{\Box} \lambda,
\end{equation}
so that $(\delta_{WZ} + \delta_L) \partial \cdot A =0$. The effect of
$\delta_L$ on all the fields in the theory is
\begin{eqnarray} \label{super}
\delta_L a &=& i\bar{\zeta} \frac{\not \! \partial}{\Box} \lambda a, 
\nonumber \\
\delta_L b &=& i\bar{\zeta} \frac{\not \! \partial}{\Box} \lambda b, 
\nonumber \\
\delta_L \psi &=& i\bar{\zeta} \frac{\not \! \partial}{\Box} \lambda \psi, 
\nonumber \\
\delta_L f &=& i\bar{\zeta} \frac{\not \! \partial}{\Box} \lambda f, 
\nonumber \\
\delta_L g &=& i\bar{\zeta} \frac{\not \! \partial}{\Box} \lambda g, 
\nonumber \\
\delta_L A_\mu &=& \partial_\mu 
\bar{\zeta} \frac{\not \! \partial}{\Box} \lambda , \nonumber \\
\delta_L \lambda &=& 0, \nonumber \\
\delta_L D &=& 0, 
\end{eqnarray}
Since $\delta_L$ is
a gauge transformation 
it leaves the gauge-invariant part of the action unchanged,
so the complete SUSY transform of SUSY QED in component notation is
\begin{equation} \label{eq:WZL}
\delta_{WZL} = \delta_{WZ} + \delta_L.
\end{equation}

This complete transform derives simple and useful identities among the 
propagators of SUSY QED~\cite{W04a}. Most notable is the identity relating
the photon and photino propagators, found thus:
\begin{eqnarray} \label{eq:Alambda}
0 &=& \langle \delta_{WZL} (A_\mu(x) \lambda(y)) \rangle \nonumber \\
&=& \langle A_\mu(x) A_\beta(y) \rangle _y\partial_\alpha
\sigma^{\beta \alpha} \zeta
- \langle \lambda(y) \bar{\lambda}(x) \rangle \gamma_\mu \zeta
+ \langle \lambda(y) \bar{\lambda}(x) \rangle 
\frac{_x\partial_\mu}{_x\!\! \not \! \partial} \zeta,
\end{eqnarray}

After converting to momentum space, the surviving members of the gauge 
multiplet have the following propagators in WZ gauge:
\bea
\langle A_\mu A_\nu \rangle (k) &=& -\frac{1}{k^2}
\left(g_{\mu \nu} - \frac{k_\mu k_\nu}{k^2}\right) \frac{1}{1+\Pi(k)}
+\xi \frac{k_\mu k_\nu}{k^4}, \label{eq:photonprop} \\
\langle \lambda \bar{\lambda} \rangle (k) &=&
\frac{-i}{A_\lambda (k) \not \! k}, \label{eq:photinoprop} \\
\langle DD\rangle (k) &=& \frac{1}{A_\lambda (k)}. \label{eq:Dprop}
\eea
where
\begin{equation} \label{eq:polarisedSUSY}
A_\lambda (k) = 1 + \Pi(k),
\end{equation} 
This relationship between the photon and photino propagators 
is what one would naively expect~\cite{KS89}.

We shall also need the identities relating the electron and selectron
propagators. One might expect that since their wavefunction
renormalisation is dependant on the gauge parameter $\xi$, 
the SUSY violation of the gauge-fixing
term would cause the electron and selectron wavefunction renormalisations 
to differ, at least nonperturbatively. However our transformations are
not violated by covariant gauge-fixing so such reasoning does not apply.
The Ward identity relating the $\psi$ and $a$ propagators is
\begin{eqnarray} \label{eq:apsi}
0 &=& \langle \delta_{WZL} (\bar{\psi}(x) a(y)) \rangle \nonumber \\
&=& -i\bar{\zeta} \langle \psi(y) \bar{\psi}(x) \rangle
+ \bar{\zeta} \langle a(y) f^*(x) \rangle 
-i\bar{\zeta}\; _x\!\!\not \! \partial \langle a(y) a^*(x) \rangle,
\end{eqnarray}
as found originally in \cite{IZ74}. 
The non-linear contribution to this and
all propagator Ward identities vanishes by the cluster decomposition
principle~\cite{OR02,W96,M93}. Taking the most general form for the electron
propagator
\begin{equation} \label{eq:eprop}
\langle \psi \psibar \rangle(p) = \frac{i}{\not \! p A(p^2) + B(p^2)},
\end{equation}
we then have
\begin{eqnarray} 
\langle a a^* \rangle (p^2) &=& \frac{A(p^2)}{p^2 A(p^2)^2 - B(p^2)^2}, 
\label{eq:aaprop} \\
\langle a f^* \rangle (p^2) = \langle f a^* \rangle (p^2) 
&=& \frac{B(p^2)}{p^2 A(p^2)^2 - B(p^2)^2}, \label{eq:afprop} \\
\langle f f^* \rangle (p^2) &=& \frac{p^2 A(p^2)}{p^2 A(p^2)^2 - B(p^2)^2},
\label{eq:ffprop}
\end{eqnarray}
where the last two propagators are given by~\cite{IZ74}
\begin{eqnarray} \label{eq:fpsi}
0 &=& \langle \delta_{WZL} (\bar{\psi}(x) f(y)) \rangle \nonumber \\
&=& \bar{\zeta} _y\not \! \partial \langle \psi(y) \bar{\psi}(x) \rangle
+ \bar{\zeta} \langle f(y) f^*(x) \rangle 
-i\bar{\zeta} _x\not \! \partial \langle f(y) a^*(x) \rangle.
\end{eqnarray}
Similarly,
\begin{eqnarray} 
\langle b b^* \rangle (p^2) &=& \frac{A(p^2)}{p^2 A(p^2)^2 - B(p^2)^2}, 
\label{eq:bbprop} \\
\langle b g^* \rangle (p^2) = \langle g b^* \rangle (p^2) 
&=& \frac{B(p^2)}{p^2 A(p^2)^2 - B(p^2)^2}, \label{eq:bgprop} \\
\langle g g^* \rangle (p^2) &=& \frac{p^2 A(p^2)}{p^2 A(p^2)^2 - B(p^2)^2}.
\label{eq:ggprop}
\end{eqnarray}
We shall use these equations in sec.~\ref{sec:vertex}. 

\section{Green's Function Ward Identities} \label{sec:Greens}
Having described our SUSY tranform and used it to restrict the form of
the propagators, we now apply the same procedure to the three-point Green's 
functions. Knowing the SUSY WIs among these will no doubt be an asset to those
computing S-matrices. As already mentioned, SUSY ST identities for these 
functions are already known~\cite{PS86,RSS01,OR02} but we are the first to use 
completely invariant WIs in SUSY QED.

We begin with an illustrative example. Consider
\begin{eqnarray}
0&=& \langle \delta_{WZL} (a(x) A_\mu(z) \psibar(y)) \rangle \nonumber \\
&=& i\zetabar \langle \psibar(y) A_\mu(z) \psi(x) \rangle 
- i\zetabar \langle (\not \! \partial a^*(y)) A_\mu(z) a(x) \rangle
+ \zetabar \langle f^*(y) A_\mu(z) a(x) \rangle \nonumber \\
&& + \zetabar \langle a(x) A_\mu(z) (a^* \not \! A)(y) \rangle
- \zetabar \gamma_\mu \langle \psibar(y) \lambda(z) a(x) \rangle
+ \zetabar  \langle \psibar(y)  
\left(\partial_\mu \frac{\not \! \partial}{\Box}\lambda(z)\right) 
a(x) \rangle.
\end{eqnarray}
These are all the non-vanishing terms that arise from our transform, 
eq.~(\ref{eq:WZL}). The term $\langle a(x) A_\mu(z) f^*(y) \rangle$ 
does not vanish but is easily
calculated using the $\langle a f^* \rangle$ propagator
(eq.~(\ref{eq:afprop})) discussed in the last section.
They are identical to those generated by the 
Wess-Zumino transforms~\cite{OR02,WZ74} except for the last term. 

The third-last term contains a non-linear contribution whose evaluation
utilises the cluster decomposition principle~\cite{OR02,W96}. A 
convenient procedure is to replace that Green's function with products of
its sub-graphs that have $a^*$ and $A_\mu$ in separate connected components.
This gives us
\begin{equation} \label{eq:cluster}
\zetabar \langle a(x) A_\mu(z) (a^* \not \! A)(y) \rangle
= \langle a(x) a^*(y) \rangle_C \gamma^\nu \langle A_\nu(y) A_\mu(z) \rangle_C.
\end{equation}
substituting this and converting all Green's functions to connected Green's
functions gives
\begin{eqnarray} \label{eq:greenexample}
0&=& -i\langle \psibar(y) A_\mu(z) \psi(x) \rangle_C 
+ i\langle (\not \! \partial a^*(y)) A_\mu(z) a(x) \rangle_C
- \langle f^*(y) A_\mu(z) a(x) \rangle_C \nonumber \\
&& 
- \langle a(x) a^*(y) \rangle_C \gamma^\nu \langle A_\nu(y) A_\mu(z) \rangle_C
+ \gamma_\mu \langle \psibar(y) \lambda(z) a(x) \rangle_C \nonumber \\
&& - \langle \psibar(y)  
\left(\partial_\mu \frac{\not \! \partial}{\Box}\lambda(z)\right) 
a(x) \rangle_C.
\end{eqnarray}
The sign changes occur because a transition from general
to connected Greens' functions causes a sign change of $(-1)^N$ for $N$-point
functions in Minkowski space~\cite{R81}.
It is a quick and easy task to check that 
eq.~(\ref{eq:greenexample}) holds at the 
bare level, something that previous calculations could not achieve without 
the introduction of ghost fields. 

Another helpful example is
\begin{eqnarray} \label{eq:greenexample2}
0 &=& \langle \delta_{WZL} (a \lambda a^*) \rangle \nonumber \\
&=& -C \langle a^*(y) \bar{\lambda}(z) \psi(x) \rangle^T_C C^{-1} \zeta
 - \langle \psibar(y) \lambda(z) a(x) \rangle_C \zeta
+i \langle a^*(y) (_z\partial_\mu A_\nu(z)) a(x) \rangle_C 
\sigma^{\nu \mu} \zeta \nonumber \\
&& - \langle \lambda(z) (\frac{\not \! \partial_x}{_x\Box} \bar{\lambda}(x))
\rangle_C \langle a(x) a^*(y) \rangle_C \zeta
+ \langle a(x) a^*(y) \rangle_C \langle \lambda \left(\bar{\lambda}(y)
\frac{\not \! \partial}{\Box}\right) \rangle_C \zeta.
\end{eqnarray}

Again, these are the same terms that one would expect from the WZ transforms,
except for the last two which come from $\delta_L$, and again the identity
vanishes simply and easily at bare level without introducing additional ghost fields.

\TABULAR[h]{|c|c||c|c|}{\hline
\raisebox{0pt}[13pt][7pt]{$\langle \delta_{WZL}(\ldots)\rangle =0$} &
\raisebox{0pt}[13pt][7pt]{SWI} &
\raisebox{0pt}[13pt][7pt]{$\langle \delta_{WZL}(\ldots)\rangle =0$} &
\raisebox{0pt}[13pt][7pt]{SWI}\\
\hline
\raisebox{0pt}[13pt][7pt]{$\langle \delta_{WZL} (a(p) a^* (q) \lambda (p-q))\rangle$}&
\raisebox{0pt}[13pt][7pt]{\ref{eq:GNalambdaa}} &
\raisebox{0pt}[13pt][7pt]{$\langle \delta_{WZL} (a(p) D(p-q) \psibar(q))\rangle$} &
\raisebox{0pt}[13pt][7pt]{\ref{eq:GNpsiDa}}\\
\hline
\raisebox{0pt}[13pt][7pt]{$\langle 
\delta_{WZL} (b(p) b^* (q) \lambda (p-q))\rangle $} &
\raisebox{0pt}[13pt][7pt]{\ref{eq:GNblambdab}} &
\raisebox{0pt}[13pt][7pt]{$\langle 
\delta_{WZL} (b(p) D(p-q) \psibar(q))\rangle $} &
\raisebox{0pt}[13pt][7pt]{\ref{eq:GNpsiDb}}\\
\hline
\raisebox{0pt}[13pt][7pt]{$\langle 
\delta_{WZL} (f(p) a^*(q) \lambda(p-q))\rangle $} &
\raisebox{0pt}[13pt][7pt]{\ref{eq:GNalambdaf}} &
\raisebox{0pt}[13pt][7pt]{$\langle 
\delta_{WZL} (f(p) D(p-q) \psibar(q))\rangle $} &
\raisebox{0pt}[13pt][7pt]{\ref{eq:GNpsiDf}}\\
\hline
\raisebox{0pt}[13pt][7pt]{$\langle 
\delta_{WZL} (g(p) b^*(q) \lambda(p-q))\rangle $} &
\raisebox{0pt}[13pt][7pt]{\ref{eq:GNblambdag}} &
\raisebox{0pt}[13pt][7pt]{$\langle 
\delta_{WZL} (\psi(p) D(p-q) g^*(q))\rangle $} &
\raisebox{0pt}[13pt][7pt]{\ref{eq:GNpsiDg}}\\
\hline
\raisebox{0pt}[13pt][7pt]{$\langle 
\delta_{WZL} (a(p) f^*(q) \lambda(p-q))\rangle $} &
\raisebox{0pt}[13pt][7pt]{\ref{eq:GNflambdaa}} &
\raisebox{0pt}[13pt][7pt]{$\langle 
\delta_{WZL} (b(p) \lambda(p-q) a^*(q))\rangle $} &
\raisebox{0pt}[13pt][7pt]{\ref{eq:GNalambdab}}\\
\hline
\raisebox{0pt}[13pt][7pt]{$\langle 
\delta_{WZL} (b(p) g^*(q) \lambda(p-q))\rangle $} &
\raisebox{0pt}[13pt][7pt]{\ref{eq:GNglambdab}} &
\raisebox{0pt}[13pt][7pt]{$\langle 
\delta_{WZL} (a(p) \lambda (p-q) b^*(q))\rangle $} &
\raisebox{0pt}[13pt][7pt]{\ref{eq:GNblambdaa}}\\
\hline
\raisebox{0pt}[13pt][7pt]{$\langle 
\delta_{WZL} (f(p) f^*(q) \lambda(p-q))\rangle $} &
\raisebox{0pt}[13pt][7pt]{\ref{eq:GNflambdaf}} &
\raisebox{0pt}[13pt][7pt]{$\langle 
\delta_{WZL} (g(p) \lambdabar (p-q) a^*(q))\rangle $} &
\raisebox{0pt}[13pt][7pt]{\ref{eq:GNalambdag}}\\
\hline
\raisebox{0pt}[13pt][7pt]{$\langle 
\delta_{WZL} (g(p) g^*(q) \lambda(p-q))\rangle $} &
\raisebox{0pt}[13pt][7pt]{\ref{eq:GNglambdag}} &
\raisebox{0pt}[13pt][7pt]{$\langle 
\delta_{WZL} (f(p) \lambdabar (p-q) b^*(q))\rangle $} &
\raisebox{0pt}[13pt][7pt]{\ref{eq:GNblambdaf}}\\
\hline
\raisebox{0pt}[13pt][7pt]{$\langle 
\delta_{WZL} (f(p) A_\mu(p-q) \psibar(q))\rangle $} &
\raisebox{0pt}[13pt][7pt]{\ref{eq:GNpsiAf}} &
\raisebox{0pt}[13pt][7pt]{$\langle 
\delta_{WZL} (a(p) \lambdabar (p-q) g^*(q))\rangle $} &
\raisebox{0pt}[13pt][7pt]{\ref{eq:GNglambdaa}}\\
\hline
\raisebox{0pt}[13pt][7pt]{$\langle 
\delta_{WZL} (g(p) A_\mu(p-q) \psibar(q))\rangle $} &
\raisebox{0pt}[13pt][7pt]{\ref{eq:GNpsiAg}} &
\raisebox{0pt}[13pt][7pt]{$\langle 
\delta_{WZL} (b(p) \lambdabar (p-q) f^*(q))\rangle $} &
\raisebox{0pt}[13pt][7pt]{\ref{eq:GNflambdab}}\\
\hline
\raisebox{0pt}[13pt][7pt]{$\langle 
\delta_{WZL} (a(p) A_\mu(p-q) \psibar(q))\rangle $} &
\raisebox{0pt}[13pt][7pt]{\ref{eq:GNpsiAa}} &
\raisebox{0pt}[13pt][7pt]{$\langle 
\delta_{WZL} (g(p) \lambdabar (p-q) f^*(q))\rangle $} &
\raisebox{0pt}[13pt][7pt]{\ref{eq:GNflambdag}}\\
\hline
\raisebox{0pt}[13pt][7pt]{$\langle 
\delta_{WZL} (b(p) A_\mu(p-q) \psibar(q))\rangle $} &
\raisebox{0pt}[13pt][7pt]{\ref{eq:GNpsiAb}} &
\raisebox{0pt}[13pt][7pt]{$\langle 
\delta_{WZL} (f(p) \lambdabar (p-q) g^*(q))\rangle $} &
\raisebox{0pt}[13pt][7pt]{\ref{eq:GNglambdaf}}\\
\hline
\multicolumn{2}{c||}{} &
\raisebox{0pt}[13pt][7pt]{$\langle 
\delta_{WZL} (\psi_\alpha (q) \psibar^\beta (p) \lambdabar^\kappa (p-q))\rangle $} &
\raisebox{0pt}[13pt][7pt]{\ref{eq:GNindices}} \\
\cline{3-4}
}{Each SWI is derived from the invariance of 
the action to the transform  $\delta_{WZL}$. The starting point of each 
SWI (indicated by its equation number) is given in this table.
\label{tab:Green}}

We give the complete set of identities in appendix \ref{app:Greens}
modulo duplications due to charge conjugation. Table \ref{tab:Green} lists
all the possible starting points for deriving identities, together with 
their corresponding equation number.

The lesson of this section is unmistakable. The three-point Green's functions
of SUSY QED are related by very simple SUSY WIs that do not require ghost
fields or STI. Conventional approaches are an adaptation of the
WZ transforms which were derived only for the gauge invariant part of
SUSY QED.

\section{Vertex Ward Identities} \label{sec:vertex}
We have demonstrated the usefulness of the WZL transform for S-matrix 
calculations, but what about renormalisation? In this case we seem to be
stuck because our transform is both non-local and 
non-linear, forbidding any calculation that
requires the Legendre transformation in any form. The conventional method
of dealing with this problem is to introduce ghost fields that couple to the
non-linear terms. While this method is valid we would like,
having come this far,
to avoid the introduction of unphysical fields. A little inspection
reveals that the propagator WIs (eqs.~(\ref{eq:eprop})-(\ref{eq:ggprop})) 
allow the propagator denominators to be factored out. Taking linear combinations
produces a comprehensive set of identities 
relating the vertices of SUSY QED to each other. We demonstrate by considering
eqs.~(\ref{eq:GNpsiAf},\ref{eq:GNpsiAa}). We transform to momentum space,
and take the combination
\begin{eqnarray} 
A(p^2) \langle \delta (\psibar(q^2) A_\mu(p-q) f(p)) \rangle 
- B(p^2) \langle \delta (\psibar(q^2) A_\mu(p-q) a(p)) \rangle,
\end{eqnarray}.

The next step is to expand each three-point Green's function in terms of its
propagators and vertex. For example, the electron-photon Green's function expands to
\begin{equation}
\langle \psibar(q) A_\mu (p-q) \psi(p) \rangle_C = \langle \psi \psibar \rangle (p)
\Gamma_{\psibar A \psi}^\nu (p,q) \langle \psi \psibar \rangle (q) 
\langle A_\nu A_\mu \rangle (p-q).
\end{equation}
Functions with scalar legs are a little more complicated. For example
\begin{equation}
\langle \psibar(q) \lambda (p-q)  a(p) \rangle_C = 
\langle \lambda \lambdabar  \rangle (p-q)
(\Gamma_{a^* \lambdabar \psi} (p,q) \langle a a^* \rangle (p)  
+ \Gamma_{f^* \lambdabar \psi} (p,q) \langle a f^* \rangle (p) )
\langle \psi \psibar \rangle (q),
\end{equation}
and
\begin{equation}
\langle \psibar (q) \lambda (p-q)  f(p) \rangle_C = 
\langle \lambda \lambdabar \rangle (p-q)
(\Gamma_{a^* \lambdabar \psi} (p,q) \langle f a^* \rangle (p) 
+ \Gamma_{f^* \lambdabar \psi} (p,q) \langle f f^* \rangle (p) )
\langle \psi \psibar \rangle (q).
\end{equation}

Substituting in this way produces the identity
\begin{eqnarray}
\lefteqn{i\sigma^{\mu \nu}(p-q)_\nu \Gamma_{\lambdabar f^* \psi}(p,q)} 
\nonumber \\
& = & \Gamma^\mu_{\psibar A \psi}(p,q) 
-i\not \! q \Gamma^\mu_{f^* A f}(p,q)
+i\Gamma^\mu_{f^* A a}(p,q) - ie\gamma^\mu \Ap2,
\end{eqnarray}
after dividing out common factors.
Similarly, the combination
\begin{eqnarray}
B(p^2) \langle \delta (\psibar(q^2) A_\mu(p-q) a(p)) \rangle 
- p^2 A(p^2) \langle \delta (\psibar(q^2) A_\mu(p-q) f(p)) \rangle,
\end{eqnarray}
gives the vertex WI
\begin{eqnarray}
\lefteqn{i\sigma^{\mu \nu}(p-q)_\nu \Gamma_{\lambdabar a^* \psi}(p,q)} 
\nonumber \\ 
& = &
i\Gamma^\mu_{a^* A a}(p,q)
- i\not \! q \Gamma^\mu_{a^* A f}(p,q) -e\gamma^\mu 
\langle \psi \psibar \rangle^{-1}(q) \nonumber
-\not \! p \Gamma^\mu_{\psibar A \psi}(p,q) \\
&& + ie\gamma^\mu \Bp2, \\
\lefteqn{i\sigma^{\mu \nu}(p-q)_\nu \Gamma_{\lambdabar b^* \psi}(p,q)} .
\end{eqnarray}
These are the only identities produced in Feynman gauge $(\xi=1)$,
In any other gauge they each produce a second identity because the terms proportional
to $(\xi-1)$ must cancel among themselves due to 
gauge invariance. Since this second identity is not found in Feynman gauge
it must not contain any additional information. This
expection is indeed fulfilled and it is straightforward to check that it
is made redundant by the Ward-Takahashi identities
(see appendix~\ref{app:WTI}) and the two-point
SUSY WIs (\ref{eq:eprop}-\ref{eq:ffprop}). 

\TABULAR[h]{|c|c|}{\hline
\raisebox{0pt}[13pt][7pt]{Combination of Green's
function Ward identities} &
\raisebox{0pt}[13pt][7pt]{Vertex SWI} \\
\hline
\raisebox{0pt}[13pt][7pt]{$q^2 \Aq2 
(p^2 \Ap2 (\mbox{\ref{eq:GNalambdaa}}) 
-\Bp2 (\mbox{\ref{eq:GNalambdaf}}))
- \Bq2 (p^2 \Ap2 (\mbox{\ref{eq:GNflambdaa}}) 
-\Bp2 (\mbox{\ref{eq:GNflambdaf}}))$}&
\raisebox{0pt}[13pt][7pt]{\ref{ephotinoa}} \\
\hline
\raisebox{0pt}[13pt][7pt]{$
q^2 \Aq2 (p^2 \Ap2 (\mbox{\ref{eq:GNblambdab}}) 
-\Bp2 (\mbox{\ref{eq:GNblambdag}}))
- \Bq2 (p^2 \Ap2 (\mbox{\ref{eq:GNglambdab}})
-\Bp2 (\mbox{\ref{eq:GNglambdag}}))$} &
\raisebox{0pt}[13pt][7pt]{\ref{ephotinob}} \\
\hline
\raisebox{0pt}[13pt][7pt]{$q^2 \Aq2 (
\Ap2 (\mbox{\ref{eq:GNalambdaf}})
-\Bp2 (\mbox{\ref{eq:GNalambdaa}}))  
- \Bq2 (\Ap2\mbox{\ref{eq:GNflambdaf}})
-\Bp2 (\mbox{\ref{eq:GNflambdaa}}))$} &
\raisebox{0pt}[13pt][7pt]{\ref{ephotinof}} \\
\hline
\raisebox{0pt}[13pt][7pt]{$
q^2 \Aq2 (\Ap2 (\mbox{\ref{eq:GNblambdag}}) 
-\Bp2 (\mbox{\ref{eq:GNblambdab}}))
- \Bq2 (\Ap2 (\mbox{\ref{eq:GNglambdag}})
-\Bp2 (\mbox{\ref{eq:GNglambdab}}))$} &
\raisebox{0pt}[13pt][7pt]{\ref{ephotinog}} \\
\hline
\raisebox{0pt}[13pt][7pt]{$
\Aq2 (\Ap2 (\mbox{\ref{eq:GNflambdaf}}) 
-\Bp2 (\mbox{\ref{eq:GNflambdaa}}))
- \Bq2 (\Ap2 (\mbox{\ref{eq:GNalambdaf}})
-\Bp2 (\mbox{\ref{eq:GNalambdaa}}))$} &
\raisebox{0pt}[13pt][7pt]{\ref{efpsi}} \\
\hline
\raisebox{0pt}[13pt][7pt]{$
\Aq2 (\Ap2 (\mbox{\ref{eq:GNglambdag}}) 
-\Bp2 (\mbox{\ref{eq:GNglambdab}}))
- \Bq2 (\Ap2 (\mbox{\ref{eq:GNblambdag}})
-\Bp2 (\mbox{\ref{eq:GNblambdab}}))$} &
\raisebox{0pt}[13pt][7pt]{\ref{egpsi}} \\
\hline
\raisebox{0pt}[13pt][7pt]{$
\Ap2 (\mbox{\ref{eq:GNpsiAf}})
-\Bp2 (\mbox{\ref{eq:GNpsiAa}})$} &
\raisebox{0pt}[13pt][7pt]{\ref{ephotone}} \\
\hline
\raisebox{0pt}[13pt][7pt]{$
\Ap2 (\mbox{\ref{eq:GNpsiAg}})
-\Bp2 (\mbox{\ref{eq:GNpsiAb}})$} &
\raisebox{0pt}[13pt][7pt]{\ref{ephotone2}} \\
\hline
\raisebox{0pt}[13pt][7pt]{$
p^2 \Ap2 (\mbox{\ref{eq:GNpsiAa}})
-\Bp2 (\mbox{\ref{eq:GNpsiAf}})$} &
\raisebox{0pt}[13pt][7pt]{\ref{apsi}} \\
\hline
\raisebox{0pt}[13pt][7pt]{$
p^2 \Ap2 (\mbox{\ref{eq:GNpsiAb}}) 
-\Bp2 (\mbox{\ref{eq:GNpsiAg}})$} &
\raisebox{0pt}[13pt][7pt]{\ref{bpsi}} \\
\hline
\raisebox{0pt}[13pt][7pt]{\ref{eq:GNindices}} &
\raisebox{0pt}[13pt][7pt]{\ref{indices}} \\
\hline
\raisebox{0pt}[13pt][7pt]{$
p^2 \Ap2 (\mbox{\ref{eq:GNpsiDa}}) 
-\Bp2 (\mbox{\ref{eq:GNpsiDf}})$} &
\raisebox{0pt}[13pt][7pt]{\ref{aDpsi}}\\
\hline
\raisebox{0pt}[13pt][7pt]{$
p^2 \Ap2 (\mbox{\ref{eq:GNpsiDb}})
-\Bp2 (\mbox{\ref{eq:GNpsiDg}})$} &
\raisebox{0pt}[13pt][7pt]{\ref{bDpsi}}\\
\hline
\raisebox{0pt}[13pt][7pt]{$
\Ap2 (\mbox{\ref{eq:GNpsiDf}})
-\Bp2 (\mbox{\ref{eq:GNpsiDa}})
(\delta \psi(y) \delta D(z) \delta f^*(p))$} &
\raisebox{0pt}[13pt][7pt]{\ref{fDpsi}}\\
\hline
\raisebox{0pt}[13pt][7pt]{$
\Ap2 (\mbox{\ref{eq:GNpsiDg}}) 
-\Bp2 (\mbox{\ref{eq:GNpsiDb}})$} &
\raisebox{0pt}[13pt][7pt]{\ref{gDpsi}}\\
\hline
\raisebox{0pt}[13pt][7pt]{$
\Aq2 (p^2 \Ap2 (\mbox{\ref{eq:GNblambdaa}}) 
-\Bp2 (\mbox{\ref{eq:GNblambdaf}}))
- \Bq2 (p^2 \Ap2 (\mbox{\ref{eq:GNglambdaa}})
-\Bp2 (\mbox{\ref{eq:GNglambdaf}}))$} &
\raisebox{0pt}[13pt][7pt]{\ref{alambdab}}\\
\hline
\raisebox{0pt}[13pt][7pt]{$
\Aq2 (p^2 \Ap2 (\mbox{\ref{eq:GNalambdab}}) 
-\Bp2 (\mbox{\ref{eq:GNalambdag}}))
- \Bq2 (p^2 \Ap2 (\mbox{\ref{eq:GNflambdab}})
-\Bp2 (\mbox{\ref{eq:GNflambdag}}))$} &
\raisebox{0pt}[13pt][7pt]{\ref{blambdaa}}\\
\hline
\raisebox{0pt}[13pt][7pt]{$
q^2 \Aq2 (p^2 \Ap2 (\mbox{\ref{eq:GNglambdaa}}) 
-\Bp2 (\mbox{\ref{eq:GNglambdaf}}))
- \Bq2 (p^2 \Ap2 (\mbox{\ref{eq:GNblambdaa}})
-\Bp2 (\mbox{\ref{eq:GNblambdaf}}))$} &
\raisebox{0pt}[13pt][7pt]{\ref{alambdag}}\\
\hline
\raisebox{0pt}[13pt][7pt]{$
\Aq2 (p^2 \Ap2 (\mbox{\ref{eq:GNflambdab}}) 
-\Bp2 (\mbox{\ref{eq:GNflambdag}}))
- \Bq2 (p^2 \Ap2 (\mbox{\ref{eq:GNalambdab}})
-\Bp2 (\mbox{\ref{eq:GNalambdag}}))$} &
\raisebox{0pt}[13pt][7pt]{\ref{blambdaf}}\\
\hline
\raisebox{0pt}[13pt][7pt]{$
q^2 \Aq2 (\Ap2 (\mbox{\ref{eq:GNalambdag}}) 
-\Bp2 (\mbox{\ref{eq:GNalambdab}}))
- \Bq2 (\Ap2 (\mbox{\ref{eq:GNflambdag}})
-\Bp2 (\mbox{\ref{eq:GNflambdab}}))$} &
\raisebox{0pt}[13pt][7pt]{\ref{glambdaa}}\\
\hline
\raisebox{0pt}[13pt][7pt]{$
q^2 \Aq2 (\Ap2 (\mbox{\ref{eq:GNblambdaf}}) 
-\Bp2 (\mbox{\ref{eq:GNblambdaa}}))
- \Bq2 (\Ap2 (\mbox{\ref{eq:GNglambdaf}})
-\Bp2 (\mbox{\ref{eq:GNglambdaa}}))$} &
\raisebox{0pt}[13pt][7pt]{\ref{flambdab}}\\
\hline
\raisebox{0pt}[13pt][7pt]{$
q^2 \Aq2 (\Ap2 (\mbox{\ref{eq:GNglambdaf}}) 
-\Bp2 (\mbox{\ref{eq:GNglambdaa}}))
- \Bq2 (\Ap2 (\mbox{\ref{eq:GNblambdaf}})
-\Bp2 (\mbox{\ref{eq:GNblambdaa}}))$} &
\raisebox{0pt}[13pt][7pt]{\ref{flambdag}}\\
\hline
\raisebox{0pt}[13pt][7pt]{$
\Aq2 (\Ap2 (\mbox{\ref{eq:GNflambdag}}) 
-\Bp2 (\mbox{\ref{eq:GNflambdab}}))
- \Bq2 (\Ap2 (\mbox{\ref{eq:GNalambdag}})
-\Bp2 (\mbox{\ref{eq:GNalambdab}}))$} &
\raisebox{0pt}[13pt][7pt]{\ref{glambdaf}}\\
\hline}
{The SWIs relating the vertices are found by taking linear combinations
of the Green's function SWIs. The corresponding combinations for each vertex
SWI (indicated by its equation number) is given in this table.
\label{tab:vertex}}

Appendix \ref{app:vertex} lists all the SUSY vertex WIs. Table \ref{tab:vertex}
lists which combination of Green's function WIs corresponds to which
vertex identity.

The derivation for some identities utilizes charge invariance, which imposes
\bea \label{eq:chgeinvar}
C\Gamma^\mu_{x^\dagger Ay}(p,q)^T C^{-1} 
= -\Gamma^\mu_{y^\dagger Ax}(-q,-p), \nonumber \\
C\Gamma_{x^\dagger Dy}(p,q)^T C^{-1} 
= -\Gamma_{y^\dagger Dx}(-q,-p) ,
\eea
where $x,y$ are members of the chiral multiplet.

The result of this section is quite unexpected as, to our knowledge, such
identities have only been derived previously using the effective action. 
Indeed, one might quite reasonably ask how this approach went
unnoticed for so long. The answer is that Ward identities were discovered
in the context of gauge symmetries where this factorisation approach simply
doesn't work. For a quick example, try to derive the original Ward-Takahashi
identity relating the electron-photon vertex to the electron propagator,
\textit{ie}.
\begin{displaymath} 
0 = \langle \delta_G (\psibar(y) A_\mu(z) \psi(x)) \rangle.
\end{displaymath}
The result is useless, $0=0$.

We finish this section by observing that the vertex Ward 
identities derived here are identical to those derived by a different method.
In that work~\cite{WB99b} we observed that the effective action could be left
in superfield form until after the Legendre transform, and then the 
WZ gauge adopted. While a Legendre transform of non-linear terms 
was not committed, one might still hold reservations about such an approach,
which in any event cannot be applied to non-Abelian theories. By deriving
the same identities in an independant way that does not use the effective 
action or Legendre transform, we vindicate that approach. We are also permitted
to profit from that earlier work which went on to find the most general
solution for the vertices, given reasonable assumptions such as charge 
conjugation invariance. We include it in this paper for convenience but 
relegate it to appendix \ref{app:solution}.

\section{Discussion} \label{sec:discussion}
Taking the SUSY transformation that leaves the entire SUSY QED action 
completely
invariant, we have derived SUSY WIs relating the three-point Green's 
functions, which are listed in appendix \ref{app:Greens}.
As was the case with propagators~\cite{W04a}, these identities 
hold at bare level without the introduction of ghost fields, in contrast to
other approaches~\cite{RSS01,OR02}. 
We then found that by utilising the SUSY 
propagator WIs, we could take linear commbinations of our Green's function
identities which revealed vertex WIs after dividing out common factors, an
approach that does not work with gauge symmetry. These identities are listed
in appendix \ref{app:vertex}.
Notably, the resulting identities are identical to those derived previously
by adopting the WZ gauge after moving to the effective action, and whose most
general solution is already known \cite{WB99b}. 
We have restated this solution in appendix \ref{app:solution}.

This paper was limited to the Abelian SUSY gauge theory. The application of
our approach to non-Abelian theories, which are clearly of greater importance,
has yet to be demonstrated. This is a topic for future work.

\appendix

\section{Green's function Ward identities} \label{app:Greens}
This appendix lists all the SUSY Green's function Ward identities of
SUSY QED modulo charge conjugation. The starting points for deriving them
are given in table \ref{tab:Green}. Note that all Green's functions 
shown here are assumed to be connected, although the subscript $_C$ has
been neglected for the sake of clarity.
\begin{eqnarray}
0 &=& C \langle a^*(q) \bar{\lambda}(p-q) \psi(p) \rangle^T C^{-1} 
 + \langle \psibar(q) \lambda(p-q) a(p) \rangle   \nonumber \\
&&
- (p-q)_\mu \langle a^*(q) A_\nu(p-q) a(p) \rangle \sigma^{\nu \mu} 
+ie \frac{(\not \! p - \not \! q)}{(p-q)^2} 
\langle \lambda \bar{\lambda} \rangle (p-q) 
\langle a a^* \rangle (q) \nonumber \\
&& -ie \langle a a^* \rangle (p)
\langle \lambda \bar{\lambda} \rangle (p-q)
\frac{(\not \! p - \not \! q)}{(p-q)^2} . \label{eq:GNalambdaa} \\
\nonumber \\
0 &=& -C\langle b^*(q) \bar{\lambda}(p-q) \psi(p) \rangle^T C^{-1}\gamma_5 
 - \langle \psibar(q) \lambda(p-q) b(p) \rangle \gamma_5  \nonumber \\
&&
-i (p-q)_\mu \langle b^*(q) A_\nu(p-q) b(p) \rangle \sigma^{\nu \mu} 
+ e\langle \lambda \bar{\lambda} \rangle (p-q)
\frac{\not \! p - \not \! q}{(p-q)^2} 
\langle b b^* \rangle (p)  \nonumber \\
&& - e\langle b b^* \rangle (q)
\langle \lambda \bar{\lambda} \rangle (p-q)
\frac{\not \! p - \not \! q}{(p-q)^2} . \label{eq:GNblambdab} \\
\nonumber \\
0 &=& iC \langle a^*(q) 
\bar{\lambda}(p-q) \psi(p) \rangle^T C^{-1} \not \! p 
 + i\langle \psibar(q) \lambda(p-q) f(p) \rangle  \nonumber \\
&&
-i (p-q)_\mu \langle a^*(q) A_\nu(p-q) f(p) \rangle \sigma^{\nu \mu} 
 - e\langle \lambda \bar{\lambda}\rangle (p-q)
\frac{(\not \! p - \not \! q)}{(p-q)^2}  
 \langle f a^* \rangle (q)  \nonumber \\
&&
+ e\langle f a^* \rangle (p) 
\langle \lambda \bar{\lambda} \rangle (p-q) 
\frac{\not \! p - \not \! q}{(p-q)^2} 
-ie \langle a a^* \rangle (p) \langle \lambda \bar{\lambda}
\rangle (p-q) . \label{eq:GNalambdaf} \\
\nonumber \\
0 &=& -C \langle b^*(q) 
\bar{\lambda}(p-q) \psi(p) \rangle^T C^{-1} \not \! p \gamma_5 
 -\langle \psibar(q) \lambda(p-q) g(p) \rangle \gamma_5  \nonumber \\
&&
-i (p-q) \langle b^*(q) A_\nu(p-q) g(p) \rangle \sigma^{\nu \mu} 
- e\langle \lambda \bar{\lambda} \rangle 
\frac{(\not \! p - \not \! q)}{(p-q)^2} 
\langle g b^* \rangle (q)  \nonumber \\
&& + e\langle g b^* \rangle (p) 
\langle \lambda \bar{\lambda} \rangle (p-q)
\frac{(\not \! p - \not \! q)}{(p-q)^2}  .
- e \langle b b^* \rangle (p) 
\langle \lambda \bar{\lambda} \rangle (p-q). 
\label{eq:GNblambdag} \\
\nonumber \\
0 &=& +i C \langle f^*(q) \bar{\lambda}(p-q) \psi(p) \rangle^T C^{-1}
-e \langle \lambda \bar{\lambda} \rangle (p-q)
\frac{(\not \! p - \not \! q)}{(p-q)^2} 
\langle a f^* \rangle (q) \nonumber \\
&&-i(p-q)_\mu \langle f^* A_\nu a \rangle \sigma^{\nu \mu}
-i \langle \psibar \lambda b \rangle \not \! q \nonumber \\
&&+e \langle \lambda \bar{\lambda} \rangle (p-q)
\frac{(\not \! p - \not \! q)}{(p-q)^2} 
\langle a f^* \rangle (p) \nonumber \\
&&- e \langle \lambda \bar{\lambda} \rangle (p-q)
\langle b b^* \rangle (p). 
\label{eq:GNflambdaa} \\
\nonumber \\
0 &=& - C \langle g^*(q) \bar{\lambda}(p-q) \psi(p) \rangle^T C^{-1} \gamma_5 
-e \langle \lambda \bar{\lambda} \rangle (p-q)
\frac{(\not \! p - \not \! q)}{(p-q)^2} 
\langle b g^* \rangle (q) \nonumber \\
&&-i(p-q)_\mu \langle g^* A_\nu b \rangle \sigma^{\nu \mu}
+ \langle \psibar \lambda b \rangle \not \! q \gamma_5 \nonumber \\
&&+e \langle \lambda \bar{\lambda} \rangle (p-q)
\frac{(\not \! p - \not \! q)}{(p-q)^2} 
\langle b g^* \rangle (p) \nonumber \\
&&+ e \langle \lambda \bar{\lambda} \rangle (p-q)
\langle b b^* \rangle (p). 
\label{eq:GNglambdab} \\
\nonumber \\
0 &=& -iC \langle f^*(q) 
\bar{\lambda}(p-q) \psi(p) \rangle^T C^{-1} \not \! p 
 + i\langle \psibar(q) \lambda(p-q) f(p) \rangle  \nonumber \\
&&
-i (p-q)_\mu \langle f^*(q) A_\nu(p-q) f(p) \rangle \sigma^{\nu \mu} 
-e \langle \lambda \bar{\lambda} \rangle (p-q)
\frac{(\not \! p - \not \! q)}{(p-q)^2} 
\langle f f^* \rangle (q)  \nonumber \\
&& +e \langle f f^* \rangle (p) 
\langle \lambda \bar{\lambda} \rangle (p-q)
\frac{(\not \! p - \not \! q)}{(p-q)^2} 
+ie \langle f a^* \rangle (p) \langle \lambda \bar{\lambda} \rangle \frac{(\not \! p - \not \! q)}{(p-q)^2}  \nonumber \\
&& -ie \langle a f^* \rangle (p) 
\langle \lambda \bar{\lambda} \rangle (p-q)
\frac{(\not \! p - \not \! q)}{(p-q)^2} . \label{eq:GNflambdaf} \\
\nonumber \\
0 &=& -C \langle g^*(q) 
\bar{\lambda}(p-q) \psi(p) \rangle^T C^{-1} \not \! p \gamma_5 
 + \langle \psibar(q) \lambda(p-q) g(p) \rangle \not \! q
\gamma_5  \nonumber \\
&&
-i (p-q)_\mu \langle g^*(q) A_\nu(p-q) g(p) \rangle \sigma^{\nu \mu} 
-e \langle \lambda \bar{\lambda}\rangle (p-q)
\frac{(\not \! p - \not \! q)}{(p-q)^2} 
\langle g g^* \rangle (q)  \nonumber \\
&& +e \langle g g^* \rangle (p) \langle \lambda \bar{\lambda} \rangle \frac{(\not \! p - \not \! q)}{(p-q)^2}   
- e \langle b g^* \rangle (q) 
\langle \lambda \bar{\lambda} \rangle (p-q) 
\nonumber \\
&&
+e \langle b g^* \rangle (p) \langle \lambda \bar{\lambda}
\rangle (p-q). 
\label{eq:GNglambdag} \\
\nonumber \\
0 &=& -i\not \! p\langle \psibar (q) A_\mu \psi (p) \rangle  
- ie \langle \psi \psibar \rangle (q) \gamma^\nu 
\langle A_\nu A_\mu \rangle (p-q)  \nonumber \\
&& + \left(\gamma_\mu -\frac{(p-q)_\mu}{\not \! p - \not \! q} \right)  
\langle \psibar(q) \lambdabar(p-q) f(p) \rangle
+ \langle f^*(q) A_\mu(p-q) f(p)\rangle  \nonumber \\
&& - \not \! q\langle a^*(q) A_\mu(p-q) f(p) \rangle
+ e \langle A_\mu A_\nu \rangle (p-q) \gamma^\nu 
\langle f a^* \rangle (p)
 \label{eq:GNpsiAf} \\
\nonumber \\
0 &=& 
\gamma_5 \not \! p \langle \psibar (q) A_\mu(p-q) \psi(p) \rangle  
+ e \gamma_5 \langle \psi \psibar \rangle (q) \gamma^\nu  
\langle A_\nu A_\mu \rangle (p-q)  \nonumber \\
&& - \left(\gamma_\mu - \frac{(p-q)_\mu}{\not \! p - \not \! q} \right)  
\langle \psibar(q) \lambdabar(p-q) g(p) \rangle
+ i\gamma_5 \langle g^*(q) A_\mu(p-q) g(p)\rangle \nonumber \\
&& -i \gamma_5 \not \! q \langle b^*(q) A_\mu(p-q) g(p) \rangle
+ e \langle A_\mu A_\nu \rangle (p-q) \gamma_5 \gamma^\nu  
\langle g b^* \rangle (q)  \label{eq:GNpsiAg} \\
\nonumber \\
0 &=& i\langle \psibar(q) A_\mu(p-q) \psi(p) \rangle 
- \left(\gamma_\mu -\frac{(p-q)_\mu}{\not \! p - \not \! q}  \right)
\langle  \psibar(q) \lambda(p-q) a(p) \rangle 
 \nonumber \\
&& + \langle f^*(q) A_\mu(p-q) a(p) \rangle 
- \not \! q \langle a^*(q) A_\mu(p-q) a(p) \rangle  \nonumber \\
&& + e\langle A_\mu A_\nu \rangle (p-q) \gamma^\nu 
\langle a a^* \rangle (p)  \label{eq:GNpsiAa} \\
\nonumber \\
0 &=& -\gamma_5 \langle \psibar(q) A_\mu(p-q) \psi(p) \rangle 
- \left(\gamma_\mu -\frac{(p-q)_\mu}{\not \! p - \not \! q} \right) 
\langle  \psibar(q) \lambda(p-q) b(p) \rangle 
 \nonumber \\
&& +i\gamma_5  \langle g^*(q) A_\mu(p-q) b(p) \rangle 
- i\gamma_5 \not \! q \langle b^*(q) A_\mu(p-q) b(p) \rangle 
\nonumber \\ &&
-ie \gamma_5  \langle A_\mu A_\nu \rangle (p-q) \gamma^\nu 
\langle b b^* \rangle (p)
\label{eq:GNpsiAb} \\
\nonumber 
\eea
\bea
0 &=& -\langle f^*(q) \lambdabar(p-q) \psi(p) \rangle_\alpha^{\;\;\gamma}
\delta_\kappa^{\;\; \beta}
-i \langle g^*(q) \lambdabar(p-q) \psi(p) \rangle_\alpha^{\;\;\gamma}
(\gamma_5)_\kappa^{\;\; \beta} \nonumber \\
&& + \langle a^*(q) \lambdabar(p-q) \psi(p) \rangle_\alpha^{\;\;\gamma}
(\not \! q)_\sigma^{\;\; \beta}
-i\langle b^*(q) \lambdabar(p-q) \psi(p) \rangle_\alpha^{\;\;\gamma}
(\not \! q \gamma_5)_\kappa^{\;\; \beta}
+ie \langle \psi \psibar \rangle_\alpha^{\;\; \beta} (p)
\delta_\kappa^{\;\; \gamma}
\nonumber \\
&&- ie\langle \psi \psibar \rangle_\alpha^{\;\; \beta} (q)
\delta_\kappa^{\;\; \gamma}
- (C)_{\kappa \alpha} (\langle \psibar(q) \lambda(p-q) f(p) \rangle^T
C^{-1})^{\beta \gamma} \nonumber \\
&&- i(\gamma_5 C)_{\kappa \alpha} (\langle \psibar(q) \lambda(p-q) g(p) \rangle^T
C^{-1})^{\beta \gamma}
- (\not \! p C)_{\kappa \alpha} \langle \psibar(q) \lambda(p-q) a(p) \rangle^T
(C^{-1})^{\beta \gamma} \nonumber \\
&&- i(\gamma_5 \not \! p C)_{\kappa \alpha} 
(\langle b^*(q) \lambdabar(p-q) \psi(p) \rangle^TC^{-1})^{\beta \gamma}
- i\langle \psibar(q) A_\nu (p-q) \psi(p) \rangle_\alpha^{\;\; \beta}
(\sigma^{\nu \mu} (p-q)_\mu)_\kappa^{\;\;\gamma} \nonumber \\
&&+ i\langle \psibar(q) D(p-q) \psi (p) \rangle_\alpha^{\;\; \beta}
(\gamma_5)_\kappa^{\;\; \gamma}. \label{eq:GNindices}
\eea
\bea
0 &=& \langle \psibar(q) D(p-q) \psi(p) \rangle 
+ \gamma_5 \langle g^*(q) D(p-q) a(p) \rangle \nonumber \\ 
&& - \not \! q \langle b^*(q) D(p-q) a(p) \rangle 
+ i\gamma_5(\not \! p - \not \! q)
\langle \psibar(q) \lambda (p-q) a(p) \rangle 
 \label{eq:GNpsiDa} \\
\nonumber \\
0 &=& \gamma_5 \langle \psibar(q) D(p-q) \psi(p) \rangle  
- \langle f^*(q) D(p-q) b(p) \rangle \nonumber \\ 
&& + \not \! q \langle b^*(q) D(p-q) a(p) \rangle 
+\gamma_5 (\not \! p - \not \! q) \langle \psibar*(q) \lambda(p-q) b(p) \rangle 
\label{eq:GNpsiDb} \\
\nonumber \\
0 &=& -i\not \! p \langle \psibar(q) D(p-q) \psi(p) \rangle  
+ i\gamma_5 \langle g^*(q) D(p-q) f(p) \rangle \nonumber \\ 
&& - i \gamma_5 \not \! q \langle b^*(q) D(p-q) f(p) \rangle 
-\gamma_5 (\not \! p - \not \! q) 
\langle \psibar(q) \lambda(p-q) f(p) \rangle 
\label{eq:GNpsiDf} \\
\nonumber \\
0 &=& \gamma_5\not \! p \langle \psibar(q) D(p-q) \psi(p) \rangle  
+ \langle f^*(q) D(p-q) g(p) \rangle \nonumber \\ 
&& - \not \! q \langle a^*(q) D(p-q) g(p) \rangle 
- \gamma_5(\not \! p - \not \! q)
\langle \psibar(q) \lambda(p-q) g(p) \rangle 
\label{eq:GNpsiDg} \\
\nonumber \\
0 &=& i\langle \psibar(q) \lambda(p-q) b(p) \rangle 
+ i\langle a^*(q) D(p-q) b(p) \rangle \gamma_5 \nonumber \\
&&
- C\langle a^*(q) \lambdabar(p-q) \psi(p) \rangle^T C^{-1}\gamma_5 
\label{eq:GNalambdab} \\
\nonumber \\
0 &=& iC\langle b^*(q) \lambdabar(p-q) \psi(p) \rangle^T C^{-1}
+ i\langle b^*(q) D(p-q) a(p) \rangle \gamma_5 \nonumber \\
&& - \langle \psibar(q) \lambda(p-q) a(p) \rangle \gamma_5 \label{eq:GNblambdaa} \\
\nonumber \\
0 &=& 
-ibar \not \! p \langle a^*(q) \lambdabar(p-q) \psi(p) \rangle
+ ie \gamma_5 \langle \lambda \lambdabar \rangle (p-q)
\langle a a^* \rangle (q) \nonumber \\ 
&& + i\gamma_5 \langle a^*(q) D(p-q) g(p) \rangle
+ i C^{-1} \langle \psibar(q) \lambda(p-q) g(p) \rangle^T C
\label{eq:GNalambdag} \\
\nonumber \\
0 &=& -i \not \! p \langle b^*(q) \lambdabar(p-q) \psi(p) \rangle
+ ie\gamma_5 \langle \lambda \lambdabar \rangle (p-q)
\langle b b^* \rangle (q) + i\gamma_5 \langle b^*(q) D(p-q) f(p) \rangle
\nonumber \\ &&
- \gamma_5 C \langle \psibar(q) \lambda(p-q) f(p) \rangle^T C^{-1}
\label{eq:GNblambdaf} \\
\nonumber \\
0 &=& i\langle g^*(q) \lambdabar(p-q) \psi(p) \rangle
+ i\gamma_5 \langle g^*(q) D(p-q) a(p) \rangle
- ie \langle a a^* \rangle (p) \langle \lambda \lambdabar \rangle (p-q)
\nonumber \\ &&
- \gamma_5 \not \! q C\langle \psibar(q) \lambda(p-q) a(p)\rangle^T
C^{-1} \label{eq:GNglambdaa} \\
\nonumber \\
0 &=& 
+i \not \! q C \langle \psibar(q) \lambda(p-q) b(p) \rangle^T C^{-1}
- \gamma_5 \langle f^*(q) \lambdabar(p-q) \psi(p) \rangle \nonumber \\ &&
-ie\gamma_5\langle bb^* \rangle (p) \langle \lambda \lambdabar \rangle (p-q)
+ i\gamma_5 \langle f^*(q) D(p-q) b(p) \rangle \label{eq:GNflambdab} \\
\nonumber \\
0 &=&
i\not \! p \langle f^*(q) \lambdabar(p-q) \psi(p) \rangle
+e\langle gb^*\rangle (p) \langle \lambda \lambdabar \rangle \nonumber \\ 
&&- \langle f^*(q) D(p-q) g(p) \rangle
-\gamma_5\not \! q C\langle \psibar(q)\lambda(p-q) g(p)\rangle^T
C^{-1} \nonumber \\
&& -e \langle a f^* \rangle (p) 
\langle \lambda \lambdabar \rangle (p-q)
\label{eq:GNflambdag} \\
\nonumber \\
0 &=&
i\gamma_5 \not \! p \langle g^*(q) \lambdabar(p-q) \psi(p) \rangle
+e\langle bg^*\rangle (p) \langle \lambda \lambdabar \rangle (p-q) \nonumber \\ 
&& + \langle g^*(q) D(p-q) f(p) \rangle
+i\not \! q C\langle \psibar(q)\lambda(p-q) f(p)\rangle^T C^{-1}
\nonumber \\
&& -e \langle a^* f \rangle (p) 
\langle \lambda \lambdabar \rangle (p-q)
\label{eq:GNglambdaf} 
\end{eqnarray}

\section{A Review of Ward-Takahashi identities} \label{app:WTI}
We give here a brief review of the Ward-Takahashi identities
These are relations between the vertices 
and the propagators that follow from gauge symmetry. 

Shown in momentum space,
the electron-photon vertex is related to the electron propagator by
\begin{equation} \label{eq:fermionWTI}
(p-q)\cdot \Gamma_{\psibar \psi} = \langle \psi \psibar \rangle^{-1}(p)
- \langle \psi \psibar \rangle^{-1}(q),
\end{equation}
where $p,q$ are the ingoing and outgoing momenta respectively.

Analogous identities hold for the scalar fields. We state them here without 
further ado.
\bea \label{eq:scalarWTI}
(p-q)\cdot \Gamma_{aAa}(p,q) = \langle a a^* \rangle^{-1} (p^2) -
\langle a a^* \rangle^{-1} (q^2), \nonumber \\
(p-q)\cdot \Gamma_{aAf}(p,q) = \langle f a^* \rangle^{-1} (p^2) -
\langle a f^* \rangle^{-1} (q^2), \nonumber \\
(p-q)\cdot \Gamma_{fAa}(p,q) = \langle a f^* \rangle^{-1} (p^2) -
\langle f a^* \rangle^{-1} (q^2), \nonumber \\
(p-q)\cdot \Gamma_{fAf}(p,q) = \langle f f^* \rangle^{-1} (p^2) -
\langle f f^* \rangle^{-1} (q^2), \nonumber \\
(p-q)\cdot \Gamma_{bAb}(p,q) = \langle b b^* \rangle^{-1} (p^2) -
\langle b b^* \rangle^{-1} (q^2), \nonumber \\
(p-q)\cdot \Gamma_{bAg}(p,q) = \langle g b^* \rangle^{-1} (p^2) -
\langle b g^* \rangle^{-1} (q^2), \nonumber \\
(p-q)\cdot \Gamma_{gAb}(p,q) = \langle b g^* \rangle^{-1} (p^2) -
\langle g b^* \rangle^{-1} (q^2), \nonumber \\
(p-q)\cdot \Gamma_{gAg}(p,q) = \langle g g^* \rangle^{-1} (p^2) -
\langle g g^* \rangle^{-1} (q^2),
\eea
where $\Gamma_{xAy}^\mu (p,q)$ is a photon
vertex with an incoming $x$ of momentum $p$ 
and an outgoing $y$ of momentum $q$. The photon therefore has momentum $p-q$.

\section{Vertex Ward identities} \label{app:vertex}
Here we list the vertex Ward identities refered to in Table \ref{tab:vertex}
\bea 
\label{ephotinoa}
\lefteqn{\gamma_\mu \Gamma^\mu_{a^* A a}(p,q)} \nonumber \\
&=& \Gamma_{\lambdabar a^* \psi}(p,q) \not \! q + e(\Bp2 - \Bq2)
+ \Gamma_{\lambdabar a^* \psi}(-q,-p) \not \! p, \\
\label{ephotinob}
\lefteqn{\gamma_\mu \Gamma^\mu_{b^* A b}(p,q)} \nonumber \\
&=& -i\Gamma_{\lambdabar b^* \psi}(p,q)\gamma_5 \not \! q - e(\Bp2 - \Bq2)
- i\Gamma_{\lambdabar b^* \psi}(-q,-p)\gamma_5 \not \! p, \\ \nonumber
\eea
\bea
\label{ephotinof}
\gamma_\mu \Gamma^\mu_{f^* A a}(p,q) + e \Ap2
&=&  \Gamma_{\lambdabar a^* \psi}(-q,-p) 
+ \Gamma_{\lambdabar f^* \psi}(p,q)\not \! q, \\
\label{ephotinog}
\gamma_\mu \Gamma^\mu_{g^* A b}(p,q) -e \Ap2
&=&  i\Gamma_{\lambdabar b^* \psi}(-q,-p) \gamma_5 
+ i\Gamma_{\lambdabar g^* \psi}(p,q) \not \! q \gamma_5, \\ \nonumber 
\eea
\bea
\label{efpsi}
\gamma_\mu \Gamma^\mu_{f^* A f}(p,q)
&=&  \Gamma_{\lambdabar f^* \psi}(-q,-p) 
- \Gamma_{\lambdabar f^* \psi}(p,q), \\
\label{egpsi}
\gamma_\mu \Gamma^\mu_{g^* A g}(p,q)
&=& i\Gamma_{\lambdabar g^* \psi}(-q,-p)\gamma_5 
- i\Gamma_{\lambdabar g^* \psi}(p,q) \gamma_5, 
\eea
\bea 
\label{ephotone}
\lefteqn{i\sigma^{\mu \nu}(p-q)_\nu \Gamma_{\lambdabar f^* \psi}(p,q)} 
\nonumber \\
& = & \Gamma^\mu_{\psibar A \psi}(p,q) 
-i\not \! q \Gamma^\mu_{f^* A f}(p,q)
+i\Gamma^\mu_{f^* A a}(p,q) - ie\gamma^\mu \Ap2, \\
\label{ephotone2}
\lefteqn{i\sigma^{\mu \nu}(p-q)_\nu \Gamma_{\lambdabar g^* \psi}(p,q)} 
\nonumber \\
& = & i\gamma_5 \Gamma^\mu_{\psibar A \psi}(p,q)
+\gamma_5 \not \! q \Gamma^\mu_{g^* A g}(p,q)
-\gamma_5 \Gamma^\mu_{g^* A b}(p,q) + e\gamma_5 \gamma^\mu \Ap2, 
\\ \nonumber 
\eea
\bea \label{apsi}
\lefteqn{i\sigma^{\mu \nu}(p-q)_\nu \Gamma_{\lambdabar a^* \psi}(p,q)} 
\nonumber \\ 
& = &
i\Gamma^\mu_{a^* A a}(p,q)
- i\not \! q \Gamma^\mu_{a^* A f}(p,q) -e\gamma^\mu 
\langle \psi \psibar \rangle^{-1}(q) \nonumber
-\not \! p \Gamma^\mu_{\psibar A \psi}(p,q) \\
&& + ie\gamma^\mu \Bp2, \\
\lefteqn{i\sigma^{\mu \nu}(p-q)_\nu \Gamma_{\lambdabar b^* \psi}(p,q)} 
\label{bpsi} \nonumber \\ & = &
-\gamma_5 \Gamma^\mu_{b^* A b}(p,q) 
+ \gamma_5 \not \! q \Gamma^\mu_{b^* A g}(p,q) \nonumber
-i\gamma_5 e\gamma^\mu \langle \psi \psibar \rangle^{-1}(q) \\
&& - i\gamma_5 \not \! p \Gamma^\mu_{\psibar A \psi}(p,q)
- e\gamma_5 \gamma^\mu \Bp2. 
\eea
These last two equations correspond to equations (4.9,4.10) 
in \cite{WB99b}, which 
contain a typo. Specifically, the argument of the inverse electron propagator
is given as $p$ rather than $q$, as we have done here.

\bea \label{indices}
0 &=& -i(\not \! q)_\kappa^{\; \; \beta}
(\Gamma_{\psibar f \lambda}(p,q))_\alpha^{\; \; \gamma}
+ (\gamma_5 \not \! q)_\kappa^{\; \; \beta}
(\Gamma_{\psibar g \lambda}(p,q))_\alpha^{\; \; \gamma} \nonumber \\
&& - i(\not \! p C)_{\alpha \kappa}
(C^{-1} \Gamma_{f^* \lambdabar \psi}(p,q))^{\gamma \beta}
+ (\not \! p \gamma_5 C)_{\alpha \kappa}
(C^{-1} \Gamma_{g^* \lambdabar \psi}(p,q))^{\gamma \beta} \nonumber \\
&&+i \delta_\kappa^{\; \; \beta}
(\Gamma_{\psibar a \lambda}(p,q))_\alpha^{\; \; \gamma}
- (\gamma_5)_\kappa^{\; \; \beta}
(\Gamma_{\psibar b \lambda}(p,q))_\alpha^{\; \; \gamma} \nonumber \\
&& + iC_{\alpha \kappa}
(C^{-1} \Gamma_{a^* \lambdabar \psi}(p,q))^{\gamma \beta}
- (\gamma_5 C)_{\alpha \kappa}
(C^{-1}_{b^* \lambdabar \psi}(p,q))^{\gamma \beta} \nonumber \\
&& + (\gamma_\nu)_\kappa^\gamma 
(\Gamma_{\psibar A \psi}^\nu (p,q))_\alpha^\beta
- (\gamma_5 (\not \! p - \not \! q))_\kappa^{\; \; \gamma}
(\Gamma_{\psibar D \psi}(p,q))_\alpha^{\; \; \beta}, 
\eea
where $C$ is the charge conjugation matrix. Note that derivation
of this idientity requires the Ward-Takahashi identity 
(\ref{eq:fermionWTI}). We obtain
\bea
0 & = &
(\not \! p - \not \! q)\gamma_5 \Tr(\Gamma_{\psibar D \psi}(p,q))
+ \gamma_\mu \Tr(\Gamma^\mu_{\psibar A}(p,q))
+ i\Gamma_{\psibar a \lambda}(p,q) \nonumber \\
&&- \gamma_5 \Gamma_{\psibar b \lambda}(p,q) 
 - i\Gamma_{\psibar a \lambda\psi}(-q,-p)
+ \gamma_5 \Gamma_{\psibar b \lambda}(-q,-p) \nonumber \\
&& - i\not \! q \Gamma_{\psibar f \lambda}(p,q) 
+ \gamma_5 \not \! q \Gamma_{\psibar g \lambda}(p,q) 
- i\not \! p \Gamma_{\psibar f \lambda}(-q,-p) \nonumber \\
&& + \gamma_5 \not \! p \Gamma_{\psibar g \lambda}(-q,-p), 
\eea
by setting $\beta = \alpha$ and summing, and
\bea \label{eDpsi}
0 & = & i\Tr(\Gamma_{\psibar a \lambda} (p,q))
- \gamma_5 \Tr(\Gamma_{\psibar b \lambda} (p,q)) 
- i\not \! q \Tr(\Gamma_{\psibar f \lambda} (p,q))  \nonumber \\
&& + \gamma_5 \not \! q \Tr(\Gamma_{\psibar g \lambda} (p,q))
- i\Gamma_{\lambdabar a^* \psi}(p,q)
+ \gamma_5 \Gamma_{\lambdabar b^* \psi}(p,q)
- i\not \! p \Gamma_{\lambdabar f^* \psi}(p,q) \nonumber \\
&& - \not \! p \gamma_5 \Gamma_{\lambdabar g^* \psi}(p,q) 
+ \gamma_\mu \Gamma^\mu_{\psibar A \psi} (p,q) 
- \gamma_5 (\not \! p - \not \! q) \Gamma_{\psibar D \psi} (p,q), \\ \nonumber
\eea
by setting $\gamma=\alpha$ and summing.
\bea \label{aDpsi}
\lefteqn{i\gamma_5 \Gamma_{\lambdabar a^* \psi}(p,q)} \nonumber \\
& = & \not \! p \Gamma_{\psibar D \psi} (p,q) 
+ \gamma_5 \Gamma_{a^* D b}(p,q)
- \gamma_5 \not \! q \Gamma_{a^* D g}(p,q), \\
\label{bDpsi} \lefteqn{i\gamma_5 \Gamma_{\lambdabar b^* \psi}(p,q)} 
\nonumber \\
& = & i\gamma_5 \not \! p \Gamma_{\psibar D \psi} (p,q) 
- i\Gamma_{b^* D a}(p,q) + i\not \! q \Gamma_{b^* D f}(p,q), 
\eea
\bea \label{fDpsi}
\lefteqn{\gamma_5 \Gamma_{f^* D b}(p,q)} \nonumber \\
& = & i\gamma_5 \Gamma_{\lambdabar f^* \psi}(p,q)
 + \gamma_5 \not \! q \Gamma_{f^* D g}(p,q) 
+ \Gamma_{\psibar D \psi} (p,q), \\
\lefteqn{\gamma_5 \Gamma_{g^* D a}(p,q)} \label{gDpsi} \nonumber \\
& = & - \Gamma_{\lambdabar g^* \psi}(p,q)
+ \gamma_5 \not \! q \Gamma_{g^* D f}(p,q)
- \Gamma_{\psibar D \psi} (p,q), 
\eea
\bea \label{alambdab}
\lefteqn{\gamma_5(\not \! p - \not \! q) \Gamma_{a^* D b}(p,q)} \nonumber \\
& = & \Gamma_{\lambdabar b^* \psi}(-q,-p) \not \! p
+ i\Gamma_{\lambdabar a^* \psi}(p,q) \not \! q \gamma_5 
+ ie\gamma_5 (\Bp2 - \Bq2), \\ \nonumber  \\
\lefteqn{\gamma_5(\not \! p - \not \! q) \Gamma_{b^* D a}(p,q)} 
\label{blambdaa} \nonumber \\
& = & i\Gamma_{\lambdabar a^* \psi}(-q,-p) \not \! p \gamma_5
+ \Gamma_{\lambdabar b^* \psi}(p,q) \not \! q
+ ie\gamma_5 (\Bp2 - \Bq2), \\ \nonumber 
\eea
\bea \label{alambdag}
\lefteqn{\gamma_5(\not \! p - \not \! q) \Gamma_{a^* D g}(p,q)} \nonumber \\ 
& = &\Gamma_{\lambdabar g^* \psi}(-q,-p) \not \! p
- i\Gamma_{\lambdabar a^* \psi}(p,q) \gamma_5 + ie\gamma_5 \Aq2, \\
\lefteqn{\gamma_5(\not \! p - \not \! q) \Gamma_{b^* D f}(p,q)} 
\label{blambdaf} \nonumber \\
& = &i\Gamma_{\lambdabar f^* \psi}(-q,-p) \not \! p \gamma_5 
- \Gamma_{\lambdabar b^* \psi}(p,q) + ie\gamma_5 \Aq2, \\
\lefteqn{\gamma_5(\not \! p - \not \! q) \Gamma_{g^* D a}(p,q)} 
\label{glambdaa} \nonumber \\ 
& = &\Gamma_{\lambdabar g^* \psi}(p,q) \not \! q
+ i\Gamma_{\lambdabar a^* \psi}(-q,-p) \gamma_5 - ie\gamma_5 \Ap2, \\
\lefteqn{\gamma_5(\not \! p - \not \! q) \Gamma_{f^* D b}(p,q)} 
\label{flambdab} \nonumber \\ 
& = &i\Gamma_{\lambdabar f^* \psi}(p,q) \not \! q \gamma_5 
+ \Gamma_{\lambdabar b^* \psi}(-q,-p) - ie\gamma_5 \Ap2, 
\eea
\bea
\gamma_5(\not \! p - \not \! q) \Gamma_{f^* D g}(p,q) 
= & \Gamma_{\lambdabar g^* \psi}(-q,-p) \label{flambdag}
- i\Gamma_{\lambdabar f^* \psi}(p,q) \gamma_5, & \\
\gamma_5(\not \! p - \not \! q) \Gamma_{g^* D f}(p,q) 
= & i\Gamma_{\lambdabar f^* \psi}(-q,-p) \gamma_5
- \Gamma_{\lambdabar g^* \psi}(p,q) \label{glambdaf}.&
\eea

\section{General Solution of SUSY Vertex Ward identities}
\label{app:solution}
Below is a solution for the SWIs of appendix \ref{app:vertex} and WTIs
of appendix \ref{app:WTI}. It is the
most general set of vertices consistent with both sets of identities and
free of kinematic singularities if one assumes charge conjugation invariance
and
\bea \label{condition}
\Gamma_{a^* A a}^\mu (p,q) = \Gamma_{b^* A b}^\mu (p,q),\nonumber \\
\Gamma_{f^* A a}^\mu (p,q) = \Gamma_{g^* A b}^\mu (p,q),\nonumber \\
\Gamma_{a^* A f}^\mu (p,q) = \Gamma_{b^* A g}^\mu (p,q),\nonumber \\
\Gamma_{f^* A f}^\mu (p,q) = \Gamma_{g^* A g}^\mu (p,q),
\eea
The proof of this is presented in \cite{WB99b}. The assumption of 
Eq.~(\ref{condition}) is true to all orders in perturbation
theory, and any nonperturbative violations of this assumption are 
restricted by the WTIs to lie completely within their transverse components.
(Note that two vertices contained typographical errors in the original paper
\cite{WB99b}. These are noted in the body of the text. We have checked that
the proof of their uniqueness is still valid.)
Our general solution is as follows: \newline
The scalar-photon vertices are
\bea
\lefteqn{\Gamma^\mu_{a^* A a}(p,q) 
= \Gamma^\mu_{b^* A b}(p,q) =
\frac{e}{p^2 - q^2}(p^2 \Ap2 - q^2 \Aq2)(p+q)^\mu} \nonumber \\
&& + [p^\mu (q^2 - p\cdot q) + q^\mu (p^2 - p\cdot q)]
T_{aa}(p^2,q^2,p\cdot q), \label{onea} \\ \nonumber \\
\lefteqn{\Gamma^\mu_{a^* A f}(p,q) = \Gamma^\mu_{b^* A g}(p,q)
= \Gamma^\mu_{f^* A a}(p,q) 
= \Gamma^\mu_{g^* A b}(p,q)} \nonumber \\
&=&\frac{-e}{p^2 - q^2}(\Bp2 - \Bq2)(p+q)^\mu \nonumber \\
&&+ [p^\mu (q^2 - p\cdot q) + q^\mu (p^2 - p\cdot q)]
T_{af}(p^2,q^2,p\cdot q), \label{oneb} \\ \nonumber \\
\lefteqn{\Gamma^\mu_{f^* A f}(p,q) 
= \Gamma^\mu_{g^* A g}(p,q) =
\frac{e}{p^2 - q^2}(\Ap2 - \Aq2)(p+q)^\mu} \nonumber \\
&&+ [p^\mu (q^2 - p\cdot q) + q^\mu (p^2 - p\cdot q)]
T_{ff}(p^2,q^2,p\cdot q),\label{onec} 
\eea
where the three functions 
$T_{aa}(p^2,q^2,p\cdot q),T_{af}(p^2,q^2,p\cdot q)$ and
$T_{ff}(p^2,q^2,p\cdot q)$, each satisfying
$T(p^2,q^2,p\cdot q) = T(q^2,p^2,p\cdot q)$, are free of
kinematic singularities and represent the only
degrees of freedom inherent in the solution. The forms (\ref{onea}) to
(\ref{onec}) are equivalent to that given by Ball and Chiu \cite{BC80} in the
context of non-SUSY scalar QED.
The photino vertices are
\bea \label{thirtysix} 
\Gamma_{\lambdabar a^* \psi}(p,q) &=& \frac{e}{p^2 - q^2}(p^2 \Ap2 - q^2 \Aq2)
+ \frac{e}{p^2 - q^2}(\Bp2 - \Bq2)\not \! q \nonumber \\
&& + \half e (p^2 - \not \! q \not \! p)T_{aa}(p^2,q^2,p\cdot q) \nonumber \\
&&+ \half e [\not \! p (p^2 - q^2) - 2\not \! q (p^2 - p\cdot q)]
T_{af}(p^2,q^2,p\cdot q), \nonumber \\
&& + \half e p^2(q^2 - \not \! p \not \! q)T_{ff}(p^2,q^2,p\cdot q), 
\eea

and
\bea \label{thirtyfive} 
\Gamma_{\lambdabar f^* \psi}(p,q) &=& \frac{-e}{p^2 - q^2}(\Ap2 - \Aq2)\not \! q
-\frac{e}{p^2 - q^2}(\Bp2 - \Bq2) \nonumber \\
&& +\half e(\not \! p - \not \! q)T_{aa}(p^2,q^2,p\cdot q) \nonumber \\
&& +  \half e(p-q)^2 T_{af}(p^2,q^2,p\cdot q) \nonumber \\
&& - \half e\not \! q(p^2 - \not \! p \not \! q)T_{ff}(p^2,q^2,p\cdot q).
\eea
The vertices of $b,g$ with the photino are 
\bea
\Gamma_{\lambdabar b^* \psi}(p,q) 
&=& i\gamma_5 \Gamma_{\lambdabar a^* \psi}(p,q), \\
\Gamma_{\lambdabar g^* \psi}(p,q) 
&=& i\gamma_5 \Gamma_{\lambdabar f^* \psi}(p,q).
\eea
The electron-photon vertex must be restricted at least to the form given by
Ball and Chiu~\cite{BC80} for non-SUSY QED. For the SUSY case we find
\bea
\Gamma^\mu_{\psibar A \psi}(p,q) &=&
\Gamma^\mu_{BC}(p,q) + \frac{ie}{p^2 - q^2}(\Ap2 - \Aq2)
	[\half T_3^\mu - T_8^\mu] \nonumber \\
&&		- \frac{ie}{p^2 - q^2}(\Bp2 - \Bq2)T_5^\mu
		+ \half ie T_{aa}(p^2,q^2,p\cdot q) T_3^\mu \nonumber \\
&& +ie T_{af}(p^2,q^2,p\cdot q)
	[\half (p-q)^2 T_5^\mu - T_1^\mu] \nonumber \\
&& + \half ie T_{ff}(p^2,q^2,p\cdot q)
	[T_2^\mu - p\cdot q T_3^\mu - (p-q)^2 T_8^\mu], 
\eea
where 
\bea
\Gamma^\mu_{BC}(p,q) &=& \half \frac{ie}{p^2 - q^2}(\not \! p + \not \! q)
(A(p^2) - A(q^2))(p+q)^\mu \nonumber \\
&& + ie\half (A(p^2) + A(q^2)) \gamma^\mu
+ \frac{ie}{p^2 - q^2}(\Bp2 - \Bq2)(p+q)^\mu, \nonumber \\
\eea
\bea \label{transverse}
T_1^\mu &=& p^\mu(q^2 - p\cdot q) + q^\mu (p^2 - p\cdot q), \\
T_2^\mu &=& (\not \! p + \not \! q)T_1^\mu, \\
T_3^\mu &=& \gamma^\mu(p-q)^2 - (\not \! p - \not \! q)(p-q)^\mu], \\
T_5^\mu &=& \sigma^{\mu \nu}(p-q)_\nu, \\
T_8^\mu &=& \half(\not \! p \not \! q \gamma^\mu - \gamma^\mu \not \! q \not \! p).
\eea

Finally there are the vertices for the $D$-boson, namely,
\bea
\lefteqn{\Gamma_{a^* D b} (p,q) = -\Gamma_{b^* D a} (p,q)} \nonumber \\
& = & \frac{ie}{p^2 - q^2}(p^2 \Ap2 - q^2 \Aq2) 
- iep\cdot q T_{a^* a}(p^2,q^2,p\cdot q) \nonumber \\
&& + ie p^2 q^2 T_{ff}(p^2,q^2,p\cdot q), \\ \nonumber \\
\lefteqn{\Gamma_{f^* D g} (p,q) = -\Gamma_{g^* D f} (p,q)} \nonumber \\
&=&\frac{ie}{p^2 - q^2}(\Ap2 - \Aq2) 
+ ie T_{a^* a}(p^2,q^2,p\cdot q) \nonumber \\
&&- ie p\cdot q T_{f^* f}(p^2,q^2,p\cdot q),
\eea
\bea
\Gamma_{g^* D a} (p,q)&=& \frac{ie}{p^2 - q^2}(\Bp2 - \Bq2) \nonumber \\
\Gamma_{g^* D a} (p,q)&=& \frac{ie}{p^2 - q^2}(\Bp2 - \Bq2) \nonumber \\
&&-ie (q^2 - p\cdot q) T_{af}(p^2,q^2,p\cdot q), \\
\Gamma_{a^* D g} (p,q) &=& \frac{-ie}{p^2 - q^2}(\Bp2 - \Bq2) \nonumber \\
&&+ie (p^2 - p\cdot q) T_{af}(p^2,q^2,p\cdot q), \\
\Gamma_{f^* D b} (p,q) &=& \frac{-ie}{p^2 - q^2}(\Bp2 - \Bq2) \nonumber \\
&&+ie (q^2 - p\cdot q) T_{af}(p^2,q^2,p\cdot q), \\
\Gamma_{b^* D f} (p,q) &=& \frac{ie}{p^2 - q^2}(\Bp2 - \Bq2) \nonumber \\
&&-ie (p^2 - p\cdot q) T_{af}(p^2,q^2,p\cdot q), 
\eea
and
\bea
\Gamma_{\psibar D \psi}(p,q) &=& 
-\half ie\gamma_5 [(\not \! p + \not \! q)T_{a^* a}(p^2,q^2,p\cdot q) 
\nonumber \\
&& -(p^2 - q^2)T_{af}(p^2,q^2,p\cdot q) \nonumber \\
&& +(\not \! q p^2 + \not \! p q^2)T_{ff}(p^2,q^2,p\cdot q)]. 
\eea
The first of these vertices has an erroneous factor of half, and the 
last vertex has an overall sign error, in \cite{WB99b}.

\providecommand{\href}[2]{#2}\begingroup\raggedright\endgroup

\end{document}